%% file: main.tex
\documentclass[10pt,conference]{IEEEtran}
\IEEEoverridecommandlockouts
\usepackage{cite}
\usepackage{amsmath,amssymb,amsfonts}
\usepackage{algorithmic}
\usepackage{graphicx}
\usepackage{textcomp}

\usepackage{subfigure}
\usepackage{amsmath}
\usepackage{url}
\usepackage{multirow}
\usepackage{floatflt}
\usepackage{listings}
\usepackage{icomma}
\usepackage{hyperref}

\usepackage{tabularx}
\usepackage{booktabs}

\usepackage[dvipsnames]{xcolor}
\usepackage{amsfonts}

\definecolor{dkgreen}{rgb}{0,0.6,0}
\definecolor{gray}{rgb}{0.5,0.5,0.5}
\definecolor{mauve}{rgb}{0.58,0,0.82}
\definecolor{shadecolor}{rgb}{0.95,0.95,0.95}

\definecolor{pblue}{rgb}{0.13,0.13,1}
\definecolor{pgreen}{rgb}{0,0.5,0}
\definecolor{pred}{rgb}{0.9,0,0}
\definecolor{pgrey}{rgb}{0.46,0.45,0.48}
\PassOptionsToPackage{usenames,dvipsnames}{color}

\lstdefinestyle{toplisting}{
  float=tp,
  floatplacement=tbp,
}

\usepackage{comment}

\usepackage{xspace}

\usepackage{csquotes}
\MakeOuterQuote{"}

\def\BibTeX{{\rm B\kern-.05em{\sc i\kern-.025em b}\kern-.08em
    T\kern-.1667em\lower.7ex\hbox{E}\kern-.125emX}}

\input{tex/macro.tex}

\begin{document}

\title{$\mu$SE: Mutation-based Evaluation of Security-focused Static Analysis Tools for Android
% thanks is needed inside title, otherwise creates a blank top page
  \thanks{
    This research is supported in part by the NSF CNS-1815336 grant.
    }
}
\author{\IEEEauthorblockN{Amit Seal Ami, Kaushal Kafle, Adwait Nadkarni, Denys Poshyvanyk}
\IEEEauthorblockA{\textit{Computer Science Department} \\
\textit{College of William \& Mary}\\
Williamsburg, VA, USA \\
\{aami@email.,
kkafle@email.,
apnadkarni@,
denys@cs.\}wm.edu
% email address or ORCID
}
\and
\IEEEauthorblockN{Kevin Moran}
\IEEEauthorblockA{\textit{Computer Science Department} \\
\textit{George Mason University}\\
Fairfax, VA, USA \\
kpmoran@gmu.edu}
}
\maketitle

\begin{abstract}
\input{tex/abstract.tex}
\end{abstract}
\noindent \textbf{Website:\hspace{0.9em}}~\texttt{\footnotesize\textbf{\url{https://muse-security-evaluation.github.io}}}
\textbf{Video {\small URL}:}~\texttt{\footnotesize\textbf{\url{https://youtu.be/Kfkzi57gYys}}}

\vspace{0.4em}
\begin{IEEEkeywords}
Security, Software, Java, Testing strategies, Security and privacy
\end{IEEEkeywords}

\input{tex/introduction.tex}
\input{tex/tech_approach.tex}
\input{tex/demonstration.tex}
\input{tex/evaluation.tex}
\input{tex/conclusion.tex}

\input{tex/acknowledgement.tex}

\bibliographystyle{./IEEEtran}
\bibliography{./bib/ref}

\end{document}

%% file: tex/macro.tex
\newcommand{\tool}{$\mu$SE\xspace}
\newcommand{\tools}{$\mu$SE's\xspace}

\newcommand{\ie}{\textit{i.e.}\xspace}
\newcommand{\eg}{\textit{e.g.}\xspace}

\newboolean{showcomments}

\setboolean{showcomments}{true}

\ifthenelse{\boolean{showcomments}}
  {\newcommand{\nb}[2]{
    \fbox{\bfseries\sffamily\scriptsize#1}
    {\sf\small$\blacktriangleright$\textit{#2}$\blacktriangleleft$}
   }
   
  }
  {\newcommand{\nb}[2]{}
   
  }

%% file: tex/abstract.tex
This demo paper presents the technical details and usage scenarios of \tool{}: a mutation-based tool for evaluating security-focused static analysis tools for Android.
Mutation testing is generally used by software practitioners to assess the robustness of a given test-suite.
However, we leverage this technique to systematically evaluate static analysis tools and uncover and document soundness issues.
\tool{}'s analysis has found 25 previously undocumented flaws in static data leak detection tools for Android.
\tool{} offers four mutation schemes, namely Reachability, Complex-reachability, TaintSink, and ScopeSink, which determine the locations of seeded mutants.
Furthermore, the user can extend \tool by customizing the API calls targeted by the mutation analysis.
\tool is also practical, as it makes use of filtering techniques based on compilation and execution criteria that reduces the number of ineffective mutations.

%% file: tex/introduction.tex
\vspace{-0.5em}
\section{Introduction}
% domain
Software practitioners use different analysis tools from both academia and industry in order to guarantee a variety of security-related application properties, such as compliance, assurance, and saftey.
Static analysis tools, also known as Static Application Security Testing (SAST) tools are often favored due to their provided soundness guarantees and performance, as dynamic analysis tools can suffer from issues such as state explosion or succumb to techniques for evading detection.

% problem
Because of the importance and impact of these SASTs, they are often evaluated by using custom benchmarks~\cite{iccbench, droidbench}.
By measuring precision and recall of detected security-related issues, the practicality and soundness of SASTs can be estimated.
However, such custom curated benchmarks may not represent the diverse programming habits of developers.
Furthermore, due to the evolving nature of APIs and frameworks, such benchmarks quickly become obsolete due to new practices and considerations.
Unintentional introduction of unsound design/implementation choices further jeopardizes the security guarantees offered by such tools by making them \textit{sound}\textit{\textbf{y}} in practice. That is, these tools may make unsound design assumptions, either knowingly or unknowingly, that can impact the accuracy of their analyses.
Given the importance of SASTs in modern development workflows, there is a need for automated techniques that can help to uncover and document potential unsound properties that affect SAST efficacy.

% what this paper is about
In this paper, we describe \tool{} (pronounced as "muse"),
a mutation-based tool for evaluating security-focused static analysis tools for the Android platform. \tool seeds mutants into applications that are analyzed by static analysis tools. Then, the tools are run on the mutated applications, and any undetected mutants are analyzed to uncover flaws in the tools.
In \tool, we define \textit{security operators} based on security goals, instead of tool-specific properties (\eg, unwanted behaviors).
As a result, entire classes of tools can be analyzed using the same security operator(s).
The effectiveness of these operators in finding unsound assumptions is enhanced through the introduction of \textit{mutation schemes}~--~strategies that define \textit{where} to apply security operators to seed mutants in applications.

% further details
\tool{} is implemented using Java and uses the Eclipse Abstract Syntax Tree (AST) framework to inject mutations.
To reduce manual effort, it applies several techniques such as checking the syntax of security operators to add necessary \texttt{try-catch} blocks and removing mutations that might lead to compilation issues.
An optional execution engine can be used with \tool{} (\eg, based on CrashScope~\cite{Moran:ICST16,Moran:ICSE17}) that helps to filter non-executable mutations specifically in Android apps.
The source code of \tool{} and its documentation are publicly available~\cite{appendix}. The full details of \tool are available in research papers~\cite{musetops}.
This paper makes the following contributions:

% \todo{finish contributions list}

\begin{itemize}
    \item We provide a description of techniques that underlie \tool{}, a tool used for evaluating security-focused static analysis tools (SASTs) for Android~\cite{musetops};
	\item We discuss how \tool can be used in practice;
    \item We describe experiments on existing SASTs for data leak detection to find previously undocumented soundness issues in tools;
    \item We provide an open-source version of \tool complete with documentation~\cite{appendix}.
\end{itemize}

%% file: tex/tech_approach.tex
\section{\tool{}: Mutation based Soundness Evaluation}\label{sec:design}
\input{img/design.tex}
We provide an overview of the design of \tool{} in Fig.~\ref{fig:design}.
The specifications related to security operators and choice of mutation scheme strategy are provided through the configuration file.
Furthermore, relevant properties, such as a software system to mutate and output paths are required to be provided.
Once these are specified, \tool{} starts by analyzing the given project to identify the application Java source files while creating a model based on the Eclipse AST framework.
Based on the selected mutation scheme, \tool{} identifies locations where security operators can be injected.
When all these locations are identified, the security operators are introduced.
Mutations are inserted and checked for syntax issues through \textit{Java Compiler Diagnostics} to \textit{kill} error-prone mutations preemptively.
Furthermore, the mutated applications can optionally be executed by an execution engine to filter source-sink type mutations which may not result in data leaks.
The analysis of SASTs is done manually by applying them to detect mutants.
Logs of results from tool analysis and inserted mutations are then compared to understand what mutations were not detected by the SASTs.
Finally, undetected mutations are analyzed to determine soundness issues in SASTs.

\subsection{Security Operators}
\input{listing/security_operator.tex}
Security Operators (SO) are manifestations of unwanted behaviors with respect to a security goal.
For example, in a data leak detection tool, a security operator defines a source of sensitive information, such as IMEI or location, which is exported to a public sink, such as a device log or storage.
As a result, the same operator can be used for all the data leak detectors (\eg FlowDroid, HornDroid, and BlueSeal~\cite{QWR18, svt+14, cgm16}).
We use the Calendar and Log APIs as example of sensitive data source and sink respectively throughout this document (Listing~\ref{lst:security_operator}).
Similarly, a security operator can be defined for evaluating SASTs that detect vulnerable SSL use, or crypto API misuse as well.
\tool{} makes this process easier by allowing the user to define custom security operators where the user needs to specify the API of source and sink as well as the variable name to be used through a configuration file.

\subsection{Mutation Schemes}
Mutation schemes define \textit{where} to apply or introduce security operators within applications.
\tool{} can be used to choose one of four pre-defined mutation schemes each of which serves a different goal.

\subsubsection{Reachability Mutation Scheme}
The reachability mutation scheme is a simple and important mutation scheme that is used for evaluating the reachability of SASTs.
The reachability scheme creates mutants by injecting security operators at every reachable location, such as methods, anonymous inner class object declarations, and class level declarations.
Because of the evolving nature of APIs and frameworks, such as newly introduced lifecycle callbacks in Android frameworks as well as interfaces and abstractions, this approach is necessary to evaluate reachability.

\subsubsection{Complex-Reachability Mutation Scheme}
\input{listing/complex_reachability.tex}
SASTs often improve runtime by preventing analysis after an arbitrary number of hops in a program call graph.
The complex reachability mutation scheme makes the path from source to sink complex by inserting pre-defined hops in between source and sink.
For our implementation of \tool{}, we defined the complex-reachability for String variable that stores sensitive information as shown in Listing~\ref{lst:complex}.
To introduce complexity, it then converts it to a String Array with garbage value \texttt{StringBuilder}, which is converted back to a String with the sensitive value.
The sensitive information is then leaked to a public sink.

\subsubsection{TaintSink Mutation Scheme}
The TaintSink mutation scheme aids in evaluating asynchronous actions in SASTs by placing the source and sink in different locations that are called asynchronously.
For example, in Android applications, \texttt{onStart()} is always called before \texttt{onResume()} lifecycle method.
This can be exploited by a malicious entity that collects and stores sensitive information in a variable in \texttt{onStart()}, and then leaks it in the \texttt{onResume()} callback.

\subsubsection{ScopeSink Mutation Scheme}
\input{listing/scopesink_scheme.tex}
The ScopeSink mutation scheme takes the concept of a visibility scope from object oriented principles into consideration when seeding mutants.
As shown in Listing~\ref{lst:scope}, \tool{} analyzes the visibility scope and determines that \texttt{childMethodA} is visible from both \texttt{ChildClass} and \texttt{ParentClass}.
As a result, it creates a variable at \texttt{ParentClass} - making it visible to both \texttt{ParentClass} and \texttt{ChildClass}.
Next, sensitive information is stored in the variable \texttt{childMethodA}.
Because this variable \texttt{dl} is visible in both \texttt{childMethodA} in \texttt{ChildClass} and \texttt{methodA} in \texttt{ParentClass}, it is leaked in both locations.

\subsection{Syntax Requirements Checker}\label{sec:syntax_requirement}
A user may define custom sources and sinks using \tool{} while specifying the fully qualified name of relevant APIs.
However, some APIs also require fulfilling syntax requirements to be used properly which the user may not be aware of or simply does not want to be concerned about.
For example, accessing system storage requires the relevant functions to be placed within \texttt{try-catch} blocks as the methods throw an exception in case of uncommon situations.
To handle this, method invocations used in both source and sink, including the method chains, are checked and put inside \texttt{try-catch} blocks as required.

\subsection{Filtering Mutants}
The number of mutants can grow quickly depending on project size.
This leads to two potential issues: placing mutations in locations that result in compilation errors and placing source and sinks in locations that are never actually executed by a given application.
As a result, the analysis of \tool's mutants becomes more difficult if mutations are not filtered.
We discuss two filtering techniques to reduce the number of mutations to further facilitate the analysis.
The first, compilability-based filter, is built-in to \tool{}, whereas the latter can be achieved through an external tool such as CrashScope~\cite{Moran:ICST16,Moran:ICSE17}.

\input{img/screenshot.tex}

\subsubsection{Compilability of Mutations}
As explained in Section~\ref{sec:syntax_requirement}, some API calls may require placement inside \texttt{try-catch} blocks in order to fulfill syntax requirements.
However, this means that such API calls enclosed in \texttt{try-catch} block will result in compilation issues if these are placed in class declaration locations (\ie outside any method), even though API calls that do not require such \texttt{try-catch} blocks to be placed at those same locations.
To handle this, and other similar corner-cases, we leverage the \texttt{javax.tools.JavaCompiler} API from the Java Development Kit (JDK) to check for compilation issues on a per case basis.
This helps remove mutations which may result in compilation issues for most, if not all, corner cases.

\subsubsection{Executability of Mutants}
Because the \texttt{TaintSink} mutation scheme distributes sources and sinks across method calls, it may result in having mutations that are compilable but that may not lie upon an executable program path when the mutated program is run.
For example, if the source is placed in \texttt{onResume} method and sink in \texttt{onStart} method, these could not be feasibly executed \textit{in sequence} during runtime.
Tools that explore as many states of a given app as possible, such as CrashScope, can be used as an execution engine to explore mutant executability.
By processing those results, mutations that are not executable can be removed.

%% file: img/design.tex
\begin{figure}[htbp]
    % \begin{center}
\vspace{-0.3cm}
        \centerline{
        \includegraphics[width=.45\textwidth]{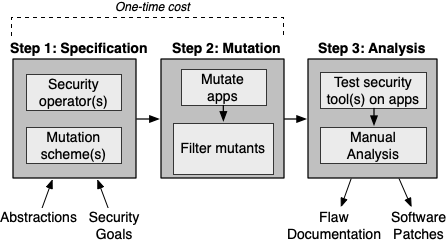}}
        \caption{\small The components and process of the \tool{}.}\label{fig:design}
    % \end{center}
\vspace{-0.4cm}
\end{figure}

%% file: listing/security_operator.tex
\begin{lstlisting}[basicstyle=\ttfamily\scriptsize,numbers=none,caption={{\small
    Security operator that injects a data leak from the Calendar API to the device log}},belowcaptionskip=-8mm,label=lst:security_operator,emph={Inject},emphstyle=\bfseries]
Inject:
String dl##
  = java.util.Calendar.getInstance().getTimeZone().
    getDisplayName();
android.util.Log.d("leak-##", dl##);
\end{lstlisting}

%% file: listing/complex_reachability.tex
\begin{lstlisting}[basicstyle=\ttfamily\scriptsize,float,numbers=none,caption={{\small
Complex Path Operator Placement
}},belowcaptionskip=-5mm,label=lst:complex,emph={Inject},emphstyle=\bfseries]
String dl0 =
    java.util.Calendar.getInstance().getTimeZone().
        getDisplayName();
String[] lr0 = new String[] {"n/a", dl0};
String dlp0 = lr0[lr0.length - 1];
android.util.Log.d("leak-0", dlp0);
\end{lstlisting}

%% file: listing/scopesink_scheme.tex
\begin{lstlisting}[basicstyle=\ttfamily\scriptsize,float,numbers=none,caption={{\small
    ScopeSink scheme based operator placement at different levels of inheritance}},belowcaptionskip=-5mm,label=lst:scope,emph={Inject},emphstyle=\bfseries]
public class ParentClass {
    String dl = "";
    int methodA(){
        android.util.Log.d("leak-0-1", dl);
        return 1; }
    class ChildClass{
        int childMethodA(){
            dl = java.util.Calendar.getInstance().
                getTimeZone().
                getDisplayName();
            android.util.Log.d("leak-0-0", dl);
            return 1; }}}
\end{lstlisting}

%% file: img/screenshot.tex
\begin{figure}[htbp]
	\vspace{-0.4cm}
    \centerline{\includegraphics[width=.52\textwidth]{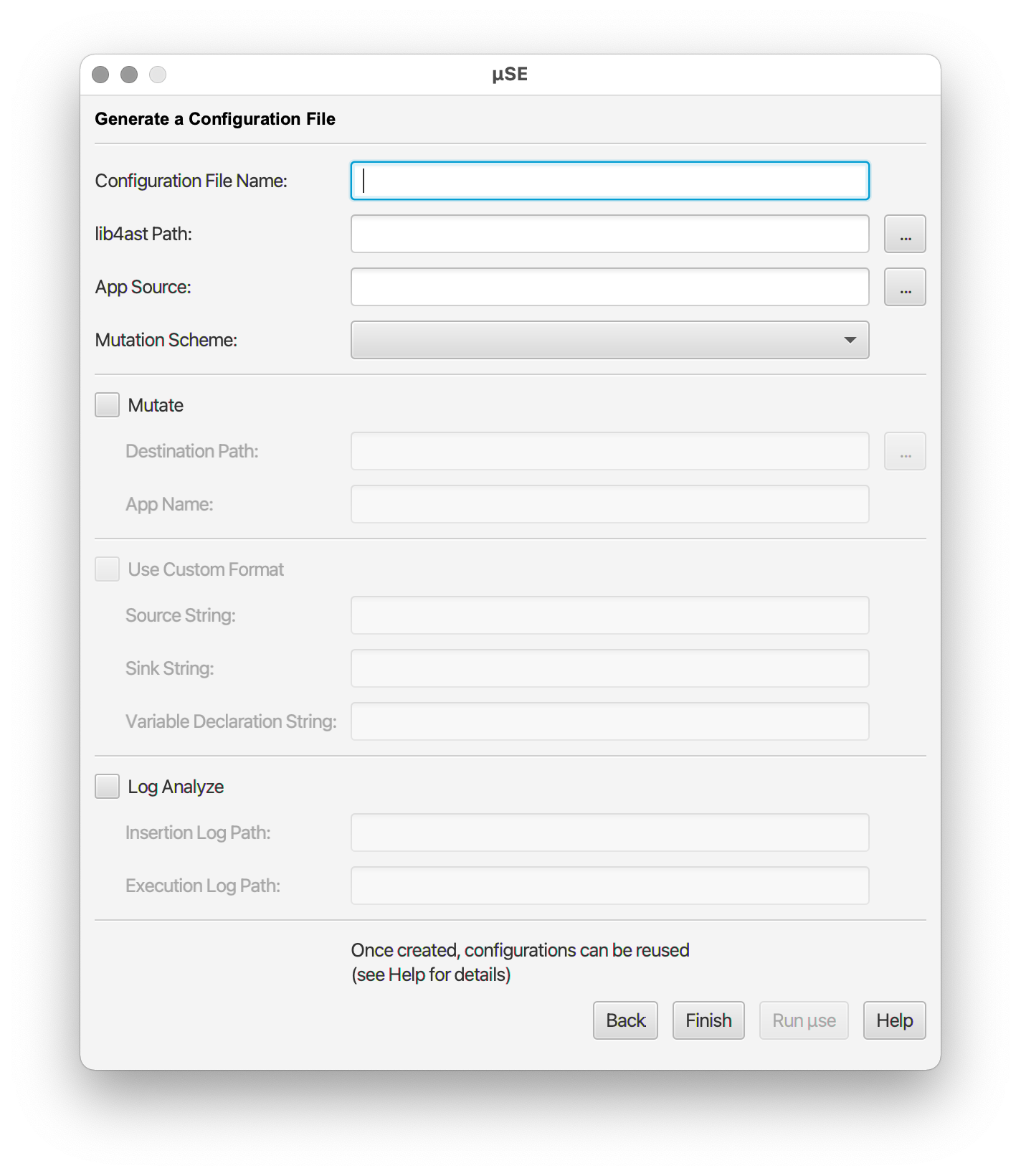}}
    \caption{Screenshot of \tool{} - Creating New Configuration File}
    \label{fig:screenshot}
\end{figure}

%% file: tex/demonstration.tex
\input{listing/configuration.tex}
\section{Usage}
\tool{} can be used through both a Command Line Interface (CLI) and a Graphical User Interface (GUI).
This allows \tool{} to be used as both standalone software as well as a component of other software.
\tool{} relies on Java LTS 11.0 for compilation and runtime.
In addition, the GUI depends on OpenJFX version 11.0.2.
Furthermore, a \texttt{JAVA\_HOME} is required to be declared as an environment variable as it is required for checking compilation issues.
To get started with mutation either through CLI or GUI, a configuration file (Listing~\ref{lst:configuration}) is required which specifies:
\begin{itemize}
    \item \textbf{lib4ast path} - contains definition libraries required for AST, such as \texttt{android.jar}
    \item \textbf{Mutation Scheme} - Choice of mutation scheme, currently available choices are Reachability, Complex-Reachability, TaintSink, and ScopeSink
    \item \textbf{App source Location} - Path of project that is to be mutated
    \item \textbf{Destination Path} - Path where mutated project will be saved
    \item \textbf{App Name} - Name of the project
\end{itemize}

\noindent By default, \tool{} uses \texttt{java.util.Calendar}, and \texttt{android.util.Log} APIs for creating source and sink related calls.
However, this can be customized by the user by specifying the variable format and involved APIs in a configuration file as shown in Listing~\ref{lst:configuration}.
\input{listing/cliuse.tex}
Because of the configuration file, using \tool{} is simple. Only the path to the configuration file is required as a parameter for the CLI as shown in Listing~\ref{lst:cli_use}.
Alternatively, the user can use the GUI as shown in Fig.~\ref{fig:screenshot} which will walk the user through different configuration settings, helping the user to save/load a configuration file for use.
\input{listing/output.tex}
When \tool{} is executed, it creates mutants while creating logs as shown in Listing~\ref{lst:output}.
Finally, the mutants are analyzed by SASTs, the results of which are analyzed with the \tools{} log as reference to identify undetected mutants for finding flaws.

%% file: listing/configuration.tex
\begin{lstlisting}[basicstyle=\ttfamily\scriptsize,float,numbers=none,caption={{\small
    Sample Configuration File for \tool{}}},belowcaptionskip=-5mm,label=lst:configuration,emph={Inject},emphstyle=\bfseries]
lib4ast: libs4ast/
appSrc: /tmp/AppFoo/src/
appName: AppFoo
output: /tmp/mutants/
operatorType: SCOPESINK
//optional customization
varDec: Type var%d = "";
source: var%d = api.method_value()
sink: api.method("value_%d-used_%d: " + var%d.toString())
\end{lstlisting}

%% file: listing/cliuse.tex
\begin{lstlisting}[basicstyle=\ttfamily\scriptsize,float,numbers=none,caption={{\small
Sample usage of \tool{} CLI }},belowcaptionskip=-5mm,label=lst:cli_use,emph={Inject},emphstyle=\bfseries]
java -jar muse.jar configuration.properties
\end{lstlisting}

%% file: listing/output.tex
\begin{lstlisting}[basicstyle=\ttfamily\scriptsize,float,numbers=none,caption={{\small
    Example Output from \tool{}}},belowcaptionskip=-5mm,label=lst:output,emph={mutation},emphstyle=\bfseries]
In file: BMIMain.java
Mutation Scheme: TAINTSINK
mutation-3-0: BMIMain.onCreate
mutation-2-0: BMIMain.onCreate
mutation-1-0: BMIMain.onCreate
mutation-0-0: BMIMain.onCreate
...
\end{lstlisting}

%% file: tex/evaluation.tex
\section{Evaluation}\label{sec:evaluation}
We used \tool{} to evaluate FlowDroid~\cite{QWR18}, a static analysis based data leak detection tool (full details are available in~\cite{musetops}) by mutating open-source Android apps.
For this, we defined a security operator that represents a data leak by collecting information related to Location and leaking it to an insecure Sink.
By systematically analyzing the results, we were able to find 13 previously documented flaws.
For example, we found that Android Fragments~\cite{android-fragments} were incorrectly modeled in FlowDroid v2.0, and the tool could not trace data leaks in Fragments. Furthermore, FlowDroid missed the \texttt{onCreate} callback of classes extending Android \texttt{SQLiteOpenHelper}. Then, we studied additional tools, namely BlueSeal~\cite{svt+14}, IccTA~\cite{lbb+15}, HornDroid~\cite{cgm16}, Argus (also known as AmanDroid~\cite{wror14}), DroidSafe~\cite{gkp+15}, and DidFail~\cite{kfb+14} to find whether the soundness issues of FlowDroid also exists for these tools.
We found that these flaws propagated to tools that were based on FlowDroid.
Furthermore, at least one of each of the FlowDroid soundness issues is applicable to each of these tools.
We later studied HornDroid and Argus using similar methodology while increasing the number of base apps for mutation (submission currently under minor acceptance review) and found 12 additional, previously undocumented flaws.
All of these flaws were communicated with the authors of these tools.
This demonstrates that \tool{} can be used to evaluate a family of Android focused SASTs for finding soundness issues.
These flaws were discoverable because of the combination of mutation augmented by the diverse practices adopted by app development practitioners.

%% file: tex/conclusion.tex
\section{Conclusion}
In this paper, we discussed the technical approach, implementation, and usage details of the mutation based soundness evaluation framework, namely  \tool{}.
Our approach helps detect previously undocumented soundness issues in Android focused SASTs~\cite{musetops} by leveraging mutation testing techniques while assimilating the diversity of real, open-source applications. As a result, developers of static analysis tools can improve soundness by applying techniques demonstrated by \tool{}.

%% file: tex/acknowledgement.tex
\section*{Acknowledgment}
The authors acknowledge the contributions from Richard Bonett for building the initial version of \tool{}, and from the following undergraduate students from William \& Mary: Liz Weech, Yang Zhang, John Clapham, Kevin Cortright, Nicholas di Mauro, Michael Foster,  Pablo Solano, Phillip Watkins, Ian Wolff, Scott Murphy, Kyle Gorham, Will Elliott \& Jeff Petit-Freres for improving  \tool{}.